# MATHEMATICAL MODELING OF BLOOD FLOW THROUGH ARTERIAL BIFURCATION

**Tahmineh Azizi[1], Robert Mugabi[2]**

[1]Institute of Computational Comparative Medicine, Department of Anatomy and Physiology, Department of Mathematics, Kansas State University, Manhattan, KS, USA

[2] Department of Vet Diagnostic & Production Animal Med, Iowa State University, Iowa, USA

**Abstract:** The blood vascular system consists of blood vessels such as arteries, arterioles, capillaries and veins that convey blood throughout the body. The pressure difference which exists between the ends of the vessels provides the living force required to the flow of blood. In this study, we have used a model that is appropriate for use with blood flow simulation applications. We study the effect of decomposition of plaques and or tumour cells on the velocity profile, pressure of the incompressible, Newtonian fluid. We consider two different cases; vessels with bifurcation and without bifurcations and we display all the results through using Commercial software COMSOL Multiphysics 5.2.

## 1. Introduction

Nanotechnology has been used a lot in the study of nanomedicine and health care related to tumor cells [1] [2] [3]. Tumor cells can adhere to the vessel walls or migrate across the endothelium to colonize in other sites [4]. Deposition of plaques of fatty materials on inner walls of blood vessels can be led to narrowing or occlusion in blood vessels and it reduces the size of lumen [5] [6]. If this procedure continues, it can produce an arterial disease. For example, in a carotid artery in which plaques have been built up, we confront with decreasing the blood flow. Atherosclerosis is a diseases that is made of due to narrows and thickens, hardens and restructures blood vessels because of build-up of plaques [7]. Treatment of this disease is by the help of Endarterectomy and Stenting which use the velocity as measured by Doppler flow in common Carotid Artery. The Doppler velocity as compared with the contra-lateral side is correlated with the degree of stenosis. For some patients, the Doppler velocity after stenting is equivalent or higher than before treatment. Some researchers believe that higher velocity measured after stenting may be due to the decreased compliance of the artery wall at stent region [7].

Atherosclerosis occurs in especial sites such as the bifurcations and flow division of the arteries and it is because of complex flow field occurring in the bifurcations of the arteries.
According to some studies, there is the correlation between atherosclerosis lesion location and low or oscillating wall shear stress [7] [8]. Moreover, many other cardiovascular diseases are relating to the flow conditions in the blood vessels. Accumulation of cholesterol and lipid particles causes hardening of the arteries and the formation of multiple plaques within the arteries. The common symptom of atherosclerotic cardiovascular disease is heart attack or sudden cardiac death [8].

The overall objective of this study is to show how velocity will change through narrow parts of the vessels due to the build-up of plaques. We considered two different cases, atherosclerotic in the bifurcations of arteries of when we have accumulation of plaques through the vessels. Commercial software COMSOL Multiphysics 5.2 has been used to solve the model and demonstrating the results. Also, we have demonstrated the pressure and velocity variations in the compliant stenotic artery.

## 2. Mathematical Background

Nowadays, the study of fluid dynamics has a significant role in fluid flow which has application inside the human body, and modelling of blood flow is a very important and difficult part in cardiovascular physics. However, models have been developed so far are very complex with three-dimensional analysis. Womersely's model makes use of Poiseuille flow and is a simplification of the Navier-Stokes equations. The Navier-Stokes equations can be used to completely model the motion of incompressible, Newtonian fluids. However, these equations are very difficult to analyse since they are non-linear, second order partial differential equations, and only in a few special cases can their exact solutions be found [9]. Poisuelli's equation may be used to develop the appropriate mathematical model for the blood pressure [10].

### 3.1 Description of model equation; Bifurcation

In this study, we assume that we have unsteady, two-dimensional, laminar, axisymmetric flow in Cartesian coordinates. Moreover, we consider that we have incompressible, Newtonian fluid, and finally we suppose that we have rigid walls with asymmetric stenosis in the mother artery. For the first case we assumed this disease in the bifurcations of arteries. See the Figure-1.

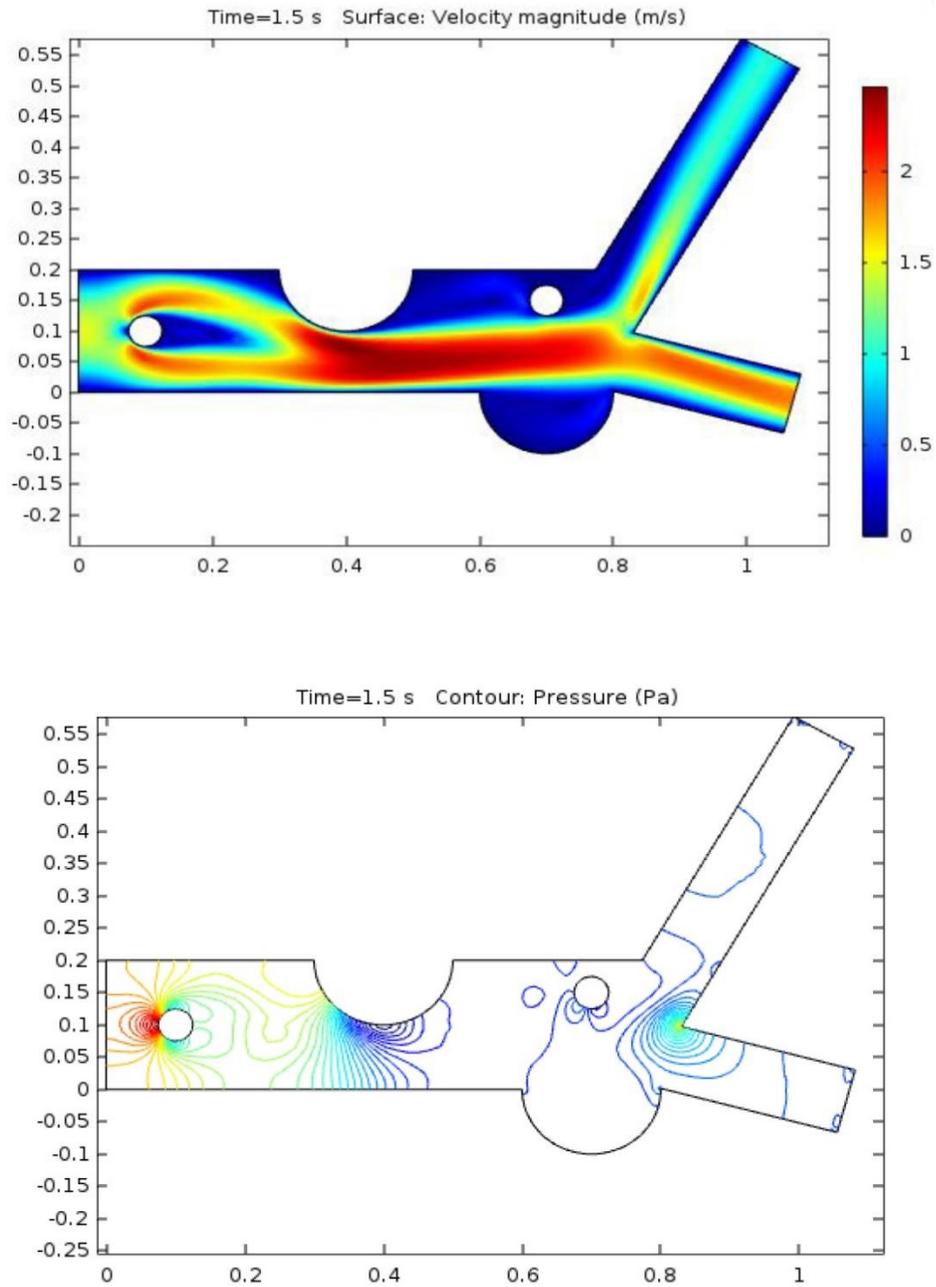

**Fig -1**: Bifurcation in vessels, Velocity magnitude and Pressure Profile

In this case, we use continuity equation as the form:

$$\frac{\partial u}{\partial x} + \frac{\partial v}{\partial y} = 0$$

Also, we use Momentum equations as the following form:

$$\rho[\frac{\partial u}{\partial t} + u\frac{\partial u}{\partial x} + v\frac{\partial u}{\partial y}] = -\frac{\partial p}{\partial x} + \mu\left(\frac{\partial^2 u}{\partial x^2} + \frac{\partial^2 u}{\partial y^2}\right) + \rho F(t)$$

$$\rho[\frac{\partial v}{\partial t} + u\frac{\partial v}{\partial x} + v\frac{\partial v}{\partial y}] = -\frac{\partial p}{\partial y} + \mu\left(\frac{\partial^2 v}{\partial x^2} + \frac{\partial^2 v}{\partial y^2}\right)$$

Where

$$F(t) = A_0 \cos(\omega_b t + \varphi)$$

$A_0$ is the amplitude, $\omega_b = 2\pi f$ is the angular velocity, $f$ is the frequency, and $\varphi$ is the lead angle of $F(t)$ with respect to heart action.

We consider the following boundary and initial conditions for the above system:

Inlet: time-dependent pressure based on pressure gradient
$$-\frac{\partial p}{\partial x} = A_0 + A_1 Cos(\omega t)$$

Outlet: Traction-free condition  
Walls: No-slip condition  
Initial Condition: average velocity $u = U_{ave}$  
With Reynolds number:
$$Re(t) = Re_{mean}[1 + 0.75\sin(2\pi\frac{t}{T}) - 0.75\cos(4\pi\frac{t}{T})]$$

Mean Reynold number=300.

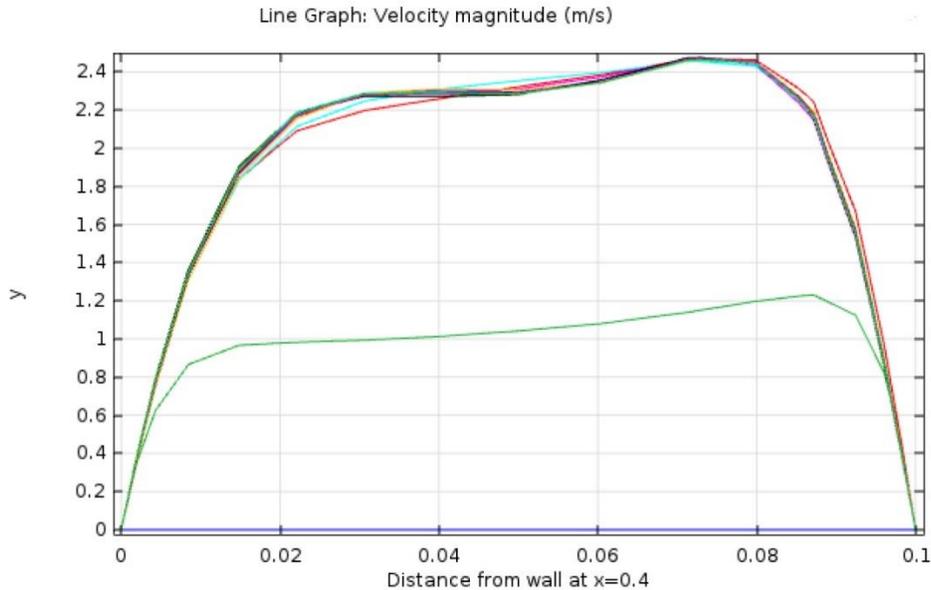

**Fig -2**: Bifurcation in vessels, Velocity magnitude and Pressure Profile

## 3.2. Flow Past a Cylinder with plaque

To study the effects of plaques on velocity and pressure, the following model analyses the unsteady, incompressible flow past a long cylinder placed in a channel at right angle to the oncoming fluid. We assume a symmetric inlet velocity profile, and the flow needs asymmetry. We consider several plaques in different parts of cylinder and we check the velocity and pressure for the fluid at those spots. The results have been demonstrated in Fig. 3-9.

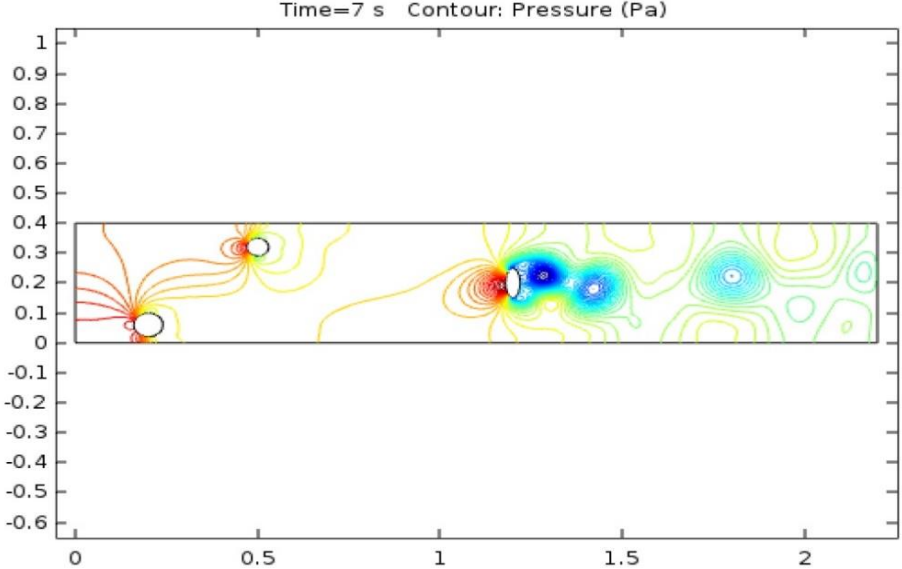

**Fig -3**: Bifurcation in vessels, Velocity magnitude and Pressure Profile

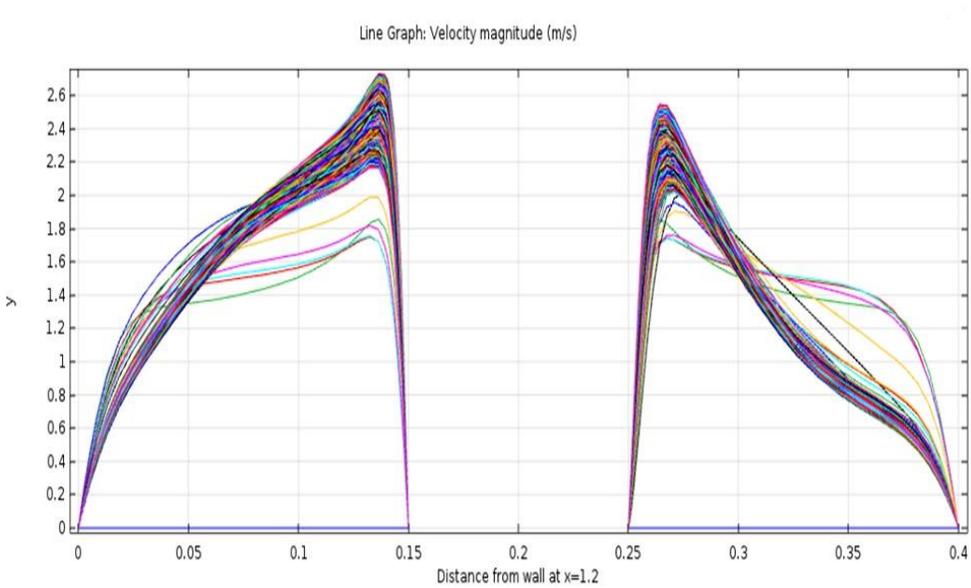

**Fig -4**: Velocity, Cylinder with plaque, $x = 1.2$

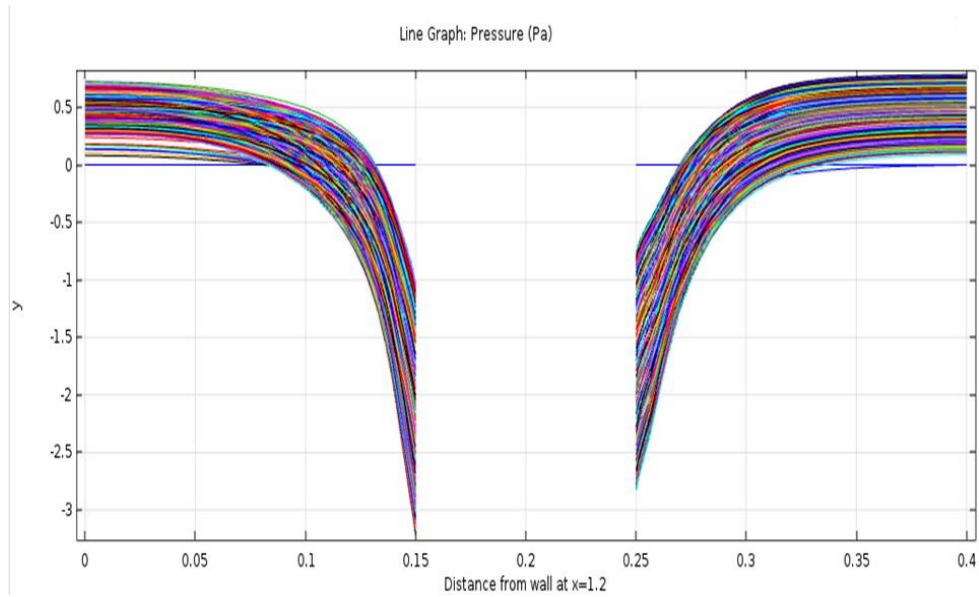

**Fig -5**: Pressure, Cylinder with plaque, $x = 1.2$

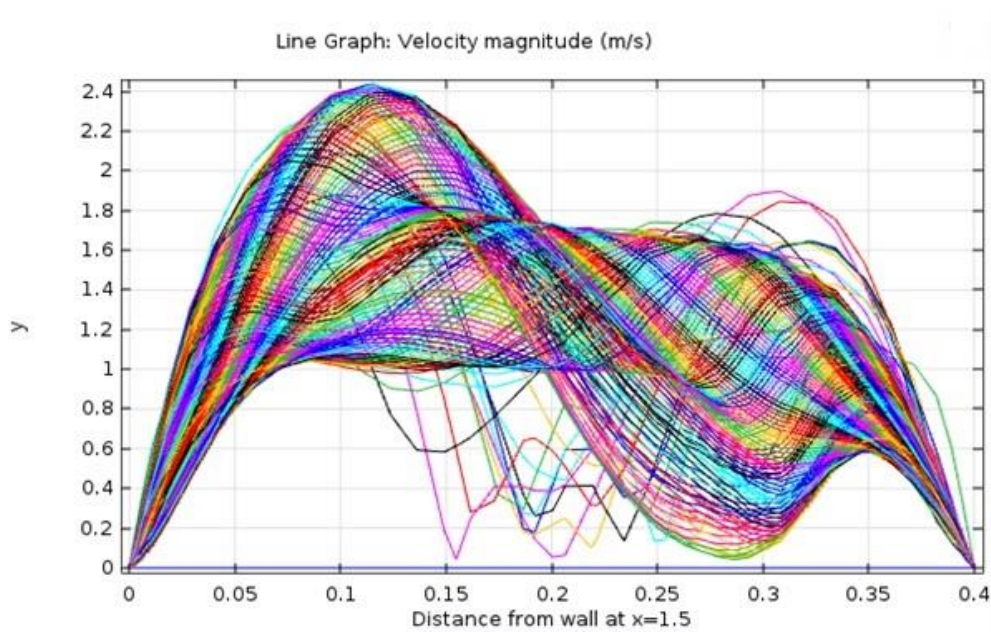

**Fig -6**: Velocity, Cylinder with plaque, $x = 1.5$

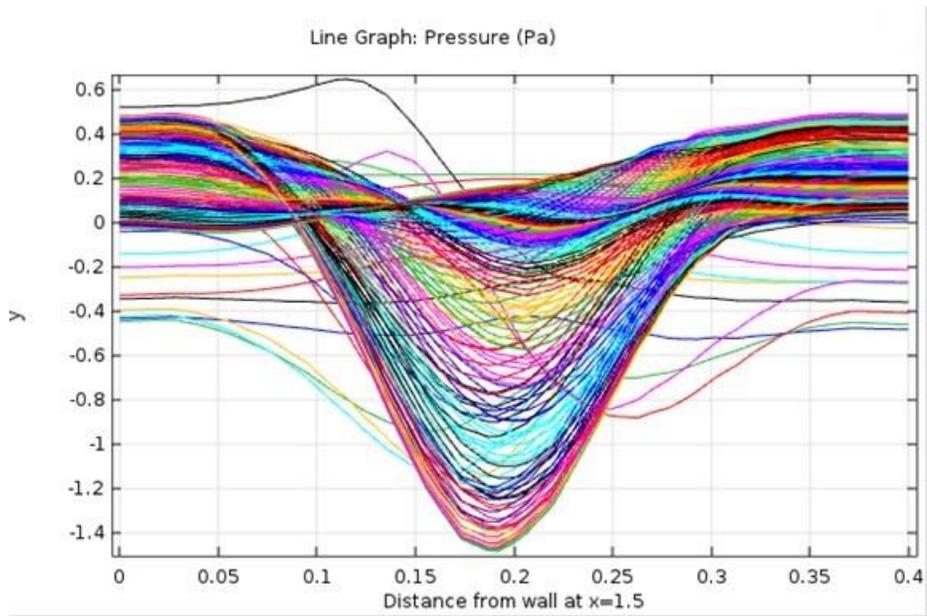

**Fig -7**: Pressure, Cylinder with plaque, $x = 1.5$

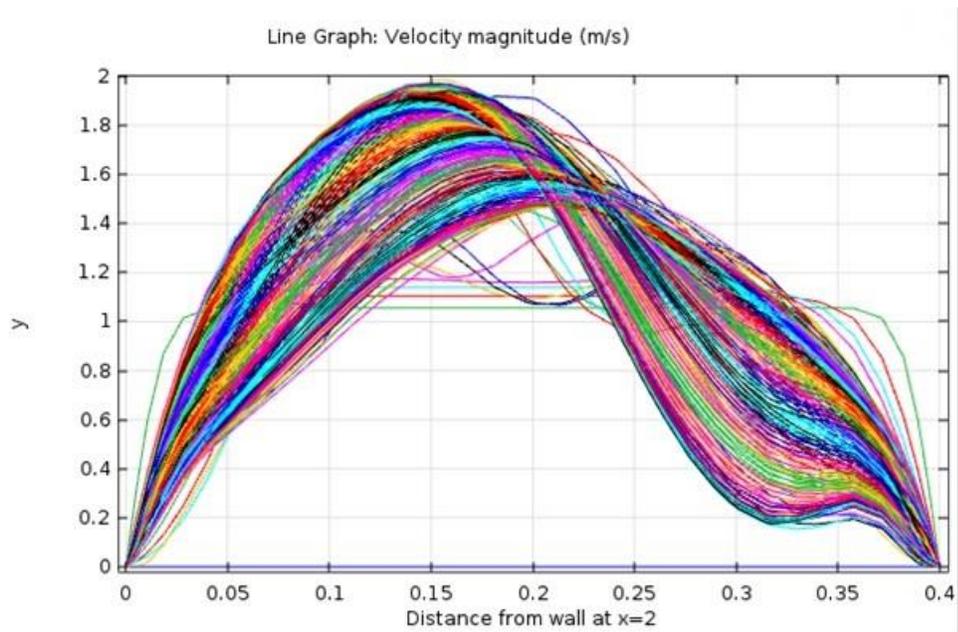

**Fig -8**: Velocity, Cylinder with plaque, $x = 2$

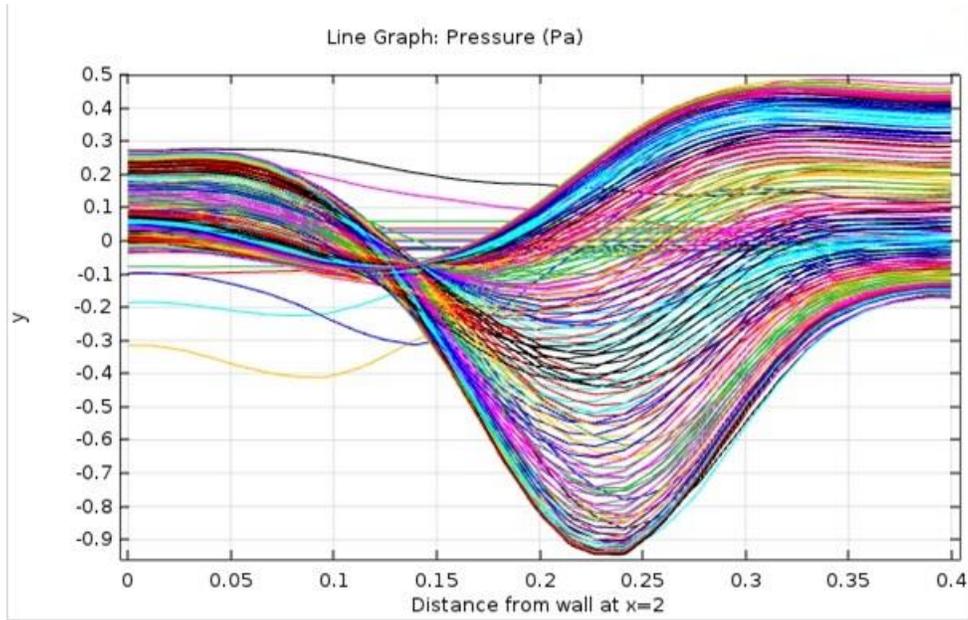

**Fig -9**: Velocity, Cylinder with plaque, $x = 2$

In Fig. 4 and Fig. 5, we can see easily how the velocity and the pressure is changing when we have decomposition of the plaques. In Fig. 6 and Fig. 7, we have demonstrated the velocity and pressure profile after the plaque in cylinder at x=1.5. Also, in Fig. 8 and Fig. 9, we have displayed the velocity profile and the pressure profile of the fluid at x=2 which is a point farther than the two last figures.

## 3. CONCLUSIONS

A mathematical model helps to describe different phenomenon in the real world. Atherosclerosis is a cardiovascular disease which happens due to narrows and thickens, hardens and restructures blood vessels because of build-up of plaques. The purpose of this paper was to present a mathematical model of the blood flow and to analysis the influence of decomposition of plaques or any tumor cells on the velocity profile and pressure profile in different kind of blood vessels. The blood flow was assumed to be laminar Newtonian, viscous and incompressible. We considered two different cases for a blood vessel and we focused on velocity profile and the pressure profile in different parts of blood vessel. Possible improvement for this study can be using other mathematical models which are more complicated to analysis the effect of plaques or tumors on different kinds of blood vessels.

# ACKNOWLEDGEMENT

This work was supported by the Institute of Computational Comparative Medicine (ICCM) and department of Mathematics of Kansas State University. With a special thanks to Dr. Majid Jaberi-Douraki for his full support.